\documentclass[superscriptaddress,prl,a4paper,twocolumn]{revtex4}%-1}

\usepackage{amsmath, amssymb} % AMS math environments

\usepackage{graphicx}

\usepackage{siunitx} % easy SI units

\usepackage{hyperref,doi}

\newcommand{\ket}[1]{\vert#1\rangle}
\newcommand{\DP}[2]{D$_{#1} \!\rightarrow \mathrm{P}_{#2}$}

\newcommand{\tableheader}[2][c]{%
  \begin{tabular}[#1]{@{}c@{}}#2\end{tabular}}
  
\long\def\symbolfootnote[#1]#2{\begingroup%
\def\thefootnote{\fnsymbol{footnote}}\footnote[#1]{#2}\endgroup}

\begin{document}

\title{Entanglement-enhanced detection of single-photon scattering events}

\author{C. Hempel}
\affiliation{Institut f\"ur Quantenoptik und Quanteninformation,
\"Osterreichische Akademie der Wissenschaften, Technikerstr.~21a, 6020 Innsbruck,
  Austria}
\affiliation{Institut f\"ur Experimentalphysik, Universit\"at Innsbruck,
  Technikerstr.~25, 6020 Innsbruck, Austria}

\author{B.~P. Lanyon}
\affiliation{Institut f\"ur Quantenoptik und Quanteninformation,
\"Osterreichische Akademie der Wissenschaften, Technikerstr.~21a, 6020 Innsbruck,
  Austria}

\author{P. Jurcevic}
\affiliation{Institut f\"ur Quantenoptik und Quanteninformation,
\"Osterreichische Akademie der Wissenschaften, Technikerstr.~21a, 6020 Innsbruck,
  Austria}
\affiliation{Institut f\"ur Experimentalphysik, Universit\"at Innsbruck,
  Technikerstr.~25, 6020 Innsbruck, Austria}

\author{R. Gerritsma}
\affiliation{Institut f\"ur Quantenoptik und Quanteninformation,
\"Osterreichische Akademie der Wissenschaften, Technikerstr.~21a, 6020 Innsbruck,
  Austria}
\affiliation{present address: QUANTUM, Institut f\"{u}r Physik, Universit\"{a}t Mainz, Staudingerweg~7, 55128 Mainz, Germany}

\author{R. Blatt}
\author{C. Roos}
\email{christian.roos@oeaw.ac.at}
\affiliation{Institut f\"ur Quantenoptik und Quanteninformation,
\"Osterreichische Akademie der Wissenschaften, Technikerstr.~21a, 6020 Innsbruck,
  Austria}
\affiliation{Institut f\"ur Experimentalphysik, Universit\"at Innsbruck,
  Technikerstr.~25, 6020 Innsbruck, Austria}

\date{July, 07 2013\footnote{uploaded later than 6 months after initial publication, according to \url{http://dx.doi.org/10.1038/nphoton.2011.218} } - journal version available at \url{http://dx.doi.org/10.1038/nphoton.2013.172}}

\maketitle

%%%%%%%%%%%%%%%%%%% Introductory paragraph %%%%%%%%%%%%%%%%%%%%%%%%%%

\textbf{
The ability to detect the interaction of light and matter at the single-particle level is becoming increasingly important for many areas of science and technology. The absorption or emission of a photon on a narrow transition of a trapped ion can be detected with near unit probability \cite{Dehmelt:1982,Leibfried:2003}, thereby enabling ultra-precise ion clocks \cite{Rosenband:2008,Chou:2010} and quantum information processing applications \cite{Haffner:2008}. Extending this sensitivity to broad transitions is challenging due to the difficulty of detecting the rapid photon scattering events in this case. Here we demonstrate a technique to detect the scattering of a single photon on a broad optical transition with high sensitivity. Our approach is to use an entangled state to amplify the tiny momentum kick an ion receives upon scattering a photon. The method should find applications in spectroscopy of atomic and molecular ions \cite{Nguyen:2011, Mur-Petit:2012, Leibfried:2012, Ding:2012} and quantum information processing.
}

%%%%%%%%%%%%%%%%%%% Main text %%%%%%%%%%%%%%%%%%%%%%%%%%

An ion in an electric trap provides an excellent system for carrying out precision spectroscopy. Laser cooling \cite{Wineland:1979}, either directly or sympathetically via an auxiliary ion \cite{Larson:1986}, minimises thermal line broadening. Long storage times allow for many repeated measurements on the same particle and collisional shifts are non-existent. 
A key requirement is to detect whether photons of a given frequency of light are scattered. For some narrow ionic transitions, corresponding to long excited-state lifetimes, the electronic configuration change associated with photon absorption or emission can be detected with near 100\% efficiency using the electron shelving technique \cite{Dehmelt:1982}. 
In the general case where electron shelving cannot be implemented directly on a `spectroscopy' ion of interest, efficient detection is possible using a co-trapped auxiliary `logic' ion following the technique of quantum logic spectroscopy \cite{Schmidt:2005}. 
Here, information about the electronic state of the spectroscopy ion is mapped via a joint vibrational mode to the electronic state of the logic ion, where it can be read out. Alternatively, state mapping can be accomplished via an off-resonantly induced optical dipole force exciting the vibrational mode conditional on the state of the spectroscopy ion \cite{Hume:2011}. For broad transitions, quantum logic spectroscopy fails due to the extremely short excited state lifetimes and resulting inability to spectrally resolve their vibrational sidebands \cite{Leibfried:2003}.

Another signature of a photon scattering event is the recoil kick an ion receives upon absorbing or emitting a photon \cite{Weiss:1993}. The size of the recoil can be characterised by the ratio of the recoil energy $E_{\mathrm{rec}}$ to the energy of a quantum of motion $h\nu$ of an harmonically trapped ion oscillating at frequency $\nu$ in the form of the dimensionless Lamb-Dicke factor $\eta{=}\sqrt{E_{\mathrm{rec}}/(h\nu)}$. For experiments on optical transitions, the Lamb Dicke factor typically satisfies  $\eta\ll 1$. For an ion cooled to its motional ground state, the probability of being promoted to the next higher vibrationally excited state by the recoil of an absorbed photon is $\eta^2$. Consequently, the method of monitoring the ground-state population is inefficient for detecting absorption events. Spectral lines from broad transitions have been reconstructed by observing changes in the fluorescence of a laser-cooled control ion via scattering of many 100's of photons\cite{Clark:2010}.

\begin{figure*}[t]
\includegraphics[scale=1]{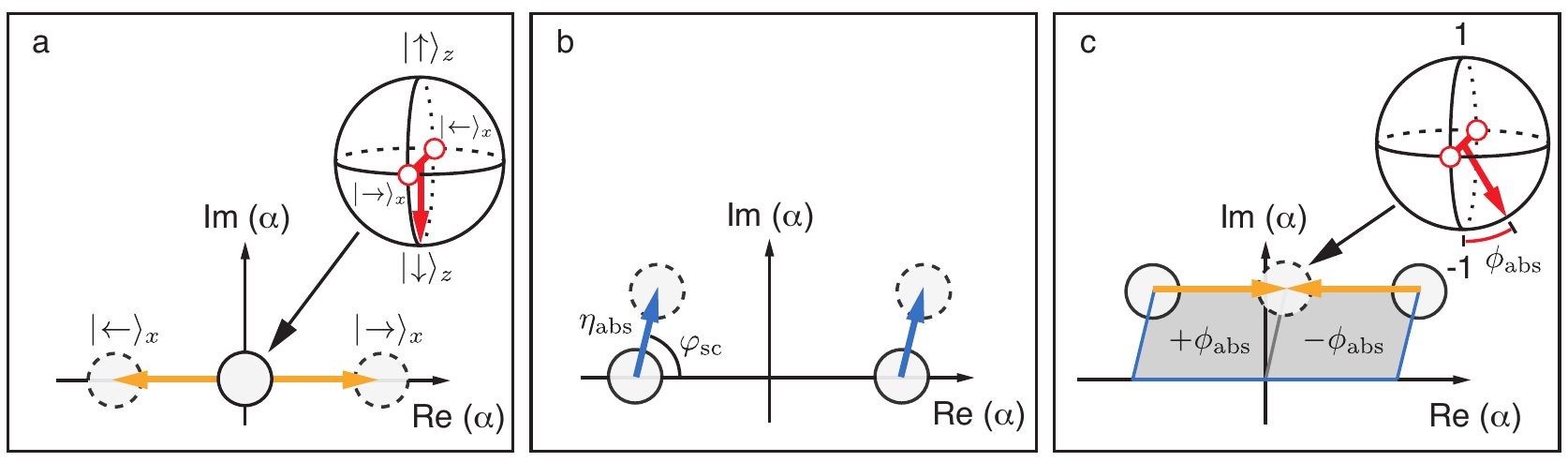}
\caption{
\textbf{Cat state spectroscopy in phase space. a} 
Schr\"{o}dinger cat state preparation. 
A qubit encoded in a logic ion, initialized in $\vert\!\downarrow\rangle_z$ (inset Bloch sphere), becomes entangled with a joint vibrational mode of the two-ion crystal formed with the co-trapped spectroscopy ion. 
\textbf{b} Absorption of a photon by the spectroscopy ion causes a displacement of size $\eta_{\mathrm{\mathrm{abs}}}$ (magnified for clarity) in a direction determined by the event's timing $\varphi_{\mathrm{sc}}$ relative to the cat-state's oscillation. 
\textbf{c} The cat-state is reinterfered, disentangling the internal state from the motion and leaving the geometric phase $\phi_\mathrm{abs}$ in the logic ion's internal state, where it can be read out via standard electron shelving.} 
\label{fig:receipe}
\end{figure*}

We now describe a technique for amplifying the recoil signal of single photons, which is independent of the particular spectroscopic ion species of interest (Fig.~\ref{fig:receipe}). As with quantum logic spectroscopy, a co-trapped logic ion encodes a two-level qubit with ground and excited states \mbox{$\ket{\!\downarrow}_z$} and \mbox{$\ket{\!\uparrow}_z$}, respectively. 
The logic qubit and a common vibrational mode of the two-ion crystal are prepared in their respective ground states.
A state-dependent force applied to this mode splits the vibrational state into two parts, each correlated with a different eigenstate in the logic qubit \cite{Poyatos:1996, Monroe:1996}.  Consequently, logic qubit and motional state become entangled into a  Schr\"{o}dinger cat state of the form \mbox{$\Psi{=}(\vert\!\rightarrow\rangle_{x}\vert\!+\!\alpha\rangle \,+\, \vert\!\leftarrow\rangle_{x}\vert\!-\!\alpha\rangle)/\sqrt{2}$}, where \mbox{$\vert\!\rightarrow\rangle_{x}$} and \mbox{$\vert\!\leftarrow\rangle_{x}$} are eigenstates of the $\sigma_x$ Pauli operator and  $\vert\!\pm\alpha\rangle$ are coherently displaced motional states. 

In the next step, spectroscopy light of a known frequency is sent into the trap. The inverse cat generation operation is then applied, thereby recombining the two vibrational components of the cat state and disentangling them from the logic qubit. If no spectroscopy photon was absorbed, the initial state of the logic qubit \mbox{$\ket{\!\downarrow}_z$} is recovered. Otherwise, the recoil of the spectroscopy ion causes both vibrational components to be displaced by $\eta_{\mathrm{abs}}$ so that after recombination the total path in phase space encloses an area. Here $\eta_{\mathrm{abs}}$ denotes the Lamb-Dicke factor of the absorbing transition. The result is a geometric phase \cite{Chaturvedi:1987} $\phi_\mathrm{abs}{=}2\alpha\eta_{\mathrm{abs}}\sin\varphi_{\mathrm{sc}}$ proportional to the cat state size~$\alpha$ and the photon recoil momentum, which leaves the final logic qubit state rotated by $\exp(i\phi_\mathrm{abs}\sigma_x)$~\cite{Turchette:2000a,Munro:2002}. Here, $\varphi_{\mathrm{sc}}{=}2\pi\nu\tau$ is the scatter phase where $\tau$ is the time delay between photon absorption and the time of largest spatial extent of $\Psi$ oscillating at frequency~$\nu$. Measurements of the logic qubit spin projections $\langle\sigma_z\rangle{=}-\cos(\phi_{\mathrm{abs}})$ or $\langle\sigma_y\rangle{=}\sin(\phi_{\mathrm{abs}})$, using standard electron shelving for example, provide information about the photon absorption process. 
In summary, the small recoil displacement of the spectroscopy ion is converted into a large geometric phase and mapped onto the logic qubit's electronic  state. 

A complete photon scattering event involves two momentum kicks -- one in absorption and one in emission.
Since the absorbed photon comes from a directed laser beam, ideally this beam points parallel to the vibrational mode supporting the cat state. The recoil direction due to the  spontaneously emitted photon is random. The effect is to add a second, random geometric phase \mbox{$\phi_{\mathrm{em}}=\tilde{\phi}_{\mathrm{em}}\cos\theta$} to the measured signal  $\langle\sigma_z\rangle{=}-\overline{\cos(\phi_\mathrm{abs}+\phi_\mathrm{em})}$, where the bar indicates averaging over the random photon emission angle $\theta$ with respect to the orientiation of the cat state in space, and $\tilde{\phi}_{\mathrm{em}}=2\alpha\eta_{\mathrm{em}}\sin\varphi_\mathrm{sc}$. For an isotropic emission pattern, this expression reduces to
\begin{equation}
\langle\sigma_z\rangle=-\cos(\phi_\mathrm{abs})\,\mathrm{sinc}(\tilde{\phi}_\mathrm{em}) \label{eq:sigma_z}
\end{equation}
\noindent and similarly $\langle\sigma_y\rangle{=}\sin(\phi_\mathrm{abs})\,\mathrm{sinc}(\tilde{\phi}_\mathrm{em})$. 

\begin{figure*}[ht!]
\includegraphics[scale=1]{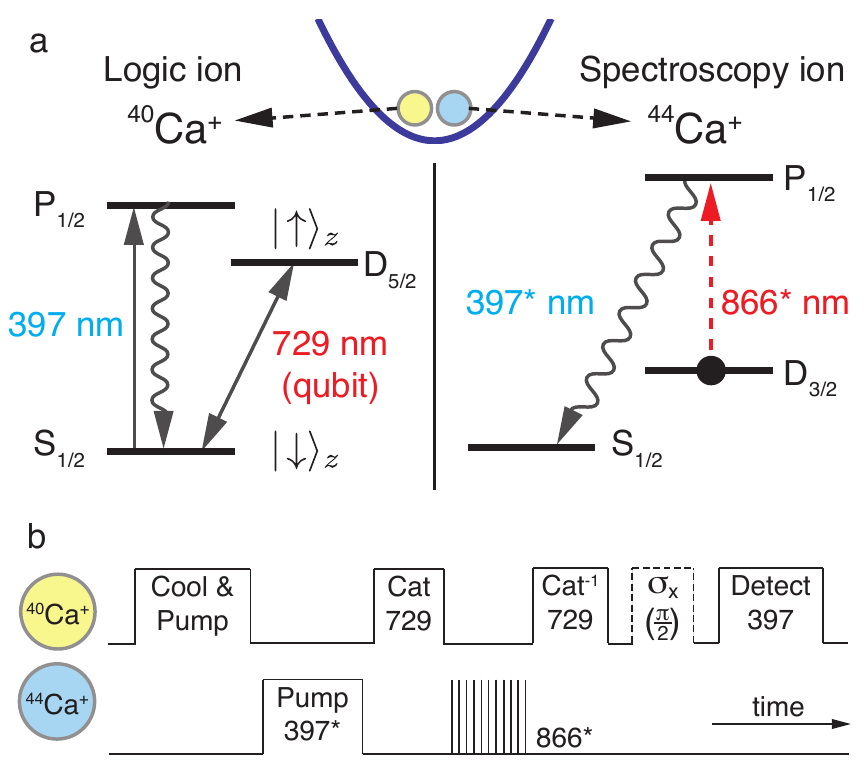}
\caption{
\textbf{Experimental details.}
\textbf{a} Relevant electronic energy levels. 
\textbf{b} Experimental laser pulse sequence.
The $^{40}$Ca$^+$ dipole transition at 397~nm is used for laser cooling, optical pumping and state detection of the optical qubit implemented on its 729~nm narrow quadrupole transition, which is also used for ground state cooling of the joint vibrational mode. The * indicates isotope shifts \cite{Lucas:2004} that allow off-resonant scattering from the respective other isotope to be safely ignored at the light intensities employed. Dedicated laser beams resonant with the $^{44}$Ca$^+$ transitions are used for initialization (397*~nm) and spectroscopy (866*~nm pulse train).}
\label{fig:levelscheme} 
\end{figure*}

\begin{figure*}[ht!]
\includegraphics[scale=1]{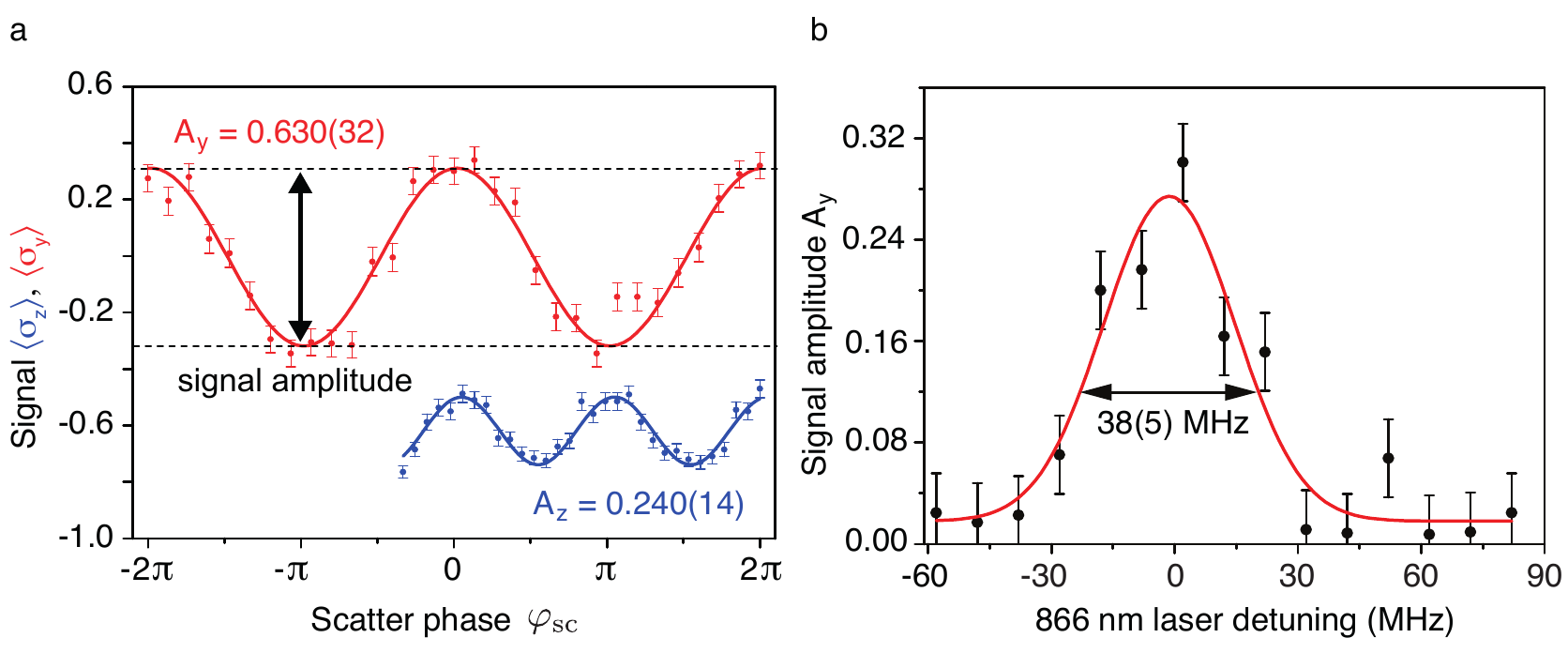}
\caption{
\textbf{Cat state spectroscopy results.}
Photon scattering signals on the broad \DP{3/2}{1/2} transition in $^{44}$Ca$^+$ for a cat state size of $\alpha = 2.9(2)$. 
\textbf{a} Interferometic fringes observed by varying $\varphi_{\mathrm{sc}}$ (the relative timing) of a single photon scattering event with respect to the cat state oscillation phase. Weighted sinusoidal fits (solid lines) yield a signal amplitude A in the two detection bases $\sigma_z$ (blue curve) and $\sigma_y$ (red curve).  To maximise the contrast for the cat state size used, the data is taken at light intensities that ensure a complete pump-out of the $D_{3/2}$ level, such that with 93.6\% probability a single (6.4\% probability $\geq 2$) photon is being scattered.
\textbf{b} Line profile of the transition taken in the $\sigma_y$ detection basis. At each spectroscopy laser frequency $A_{\mathrm{y}}$ is measured by alternating between maximum and minimum of the fringe pattern. To avoid power broadening, the light level was reduced significantly, yielding a smaller signal amplitude as, even on resonance, not in all experiments a photon is being scattered (see Supplementary information). All error bars are calculated from quantum projection noise.}
\label{fig:oscillations}
\end{figure*}

In our demonstration we confine a mixed-species two-ion crystal in a linear Paul trap (see Methods). The isotopes $^{40}$Ca$^+$ and $^{44}$Ca$^+$ are employed as logic and spectroscopy ion, respectively (Fig.~\ref{fig:levelscheme}a). 
The $^{40}$Ca$^+$ provides sympathetic Doppler and vibrational ground state cooling of the lowest frequency axial vibrational mode at \mbox{$\nu=1.199$~MHz}, a narrow electronic transition in which to encode a qubit ($\mathrm{S}_{1/2}{\,\rightarrow\,}\mathrm{D}_{5/2}$) and qubit state readout via electron shelving. We optically pump $^{44}$Ca$^+$ to its metastable $\mathrm{D}_{3/2}$ state in order to perform spectroscopy on the strong $\mathrm{D}_{3/2}{\,\rightarrow\,}\mathrm{P}_{1/2}$ open dipole transition. Absorption of a single infrared photon populates the $\mathrm{P}_{1/2}$ state (lifetime of 7.1~ns~\cite{Jin:1993}) which -  with $93.6\%$ probability\cite{Ramm:2013} - will decay to the $\mathrm{S_{1/2}}$ ground state under emission of a single blue photon.
The experimental sequence is presented in Fig.~\ref{fig:levelscheme}b.
 A state-dependent force \cite{Haljan:2005} is realised using a bichromatic laser field resonant with the red~(${-}\nu$) and blue~(${+}\nu$) vibrational sidebands of the logic qubit. 
A train of short ($\sim$60~ns) spectroscopy pulses, separated by $1/\nu$ is shifted in time by $\tau$  to vary $\varphi_\mathrm{sc}$ in order to retrieve the fringe pattern shown in Fig.~\ref{fig:oscillations}.a for a cat state size of $\alpha = 2.9(2)$. Precise timing of photon absorption is critical for the measurement of $\langle\sigma_y\rangle$. A discussion of timing issues and data for other values of $\alpha$ is presented in the Supplementary Sections S2 and S5. Since the signal $\langle\sigma_z\rangle$ (blue curve) is insensitive to the sign of the geometric phase shift, the fringe period is equal to half the motional period. Conversely, the signal $\langle\sigma_y\rangle$ (red curve) is sign-sensitive and therefore oscillates around 0 with a period equal to the motional period $1/\nu$. Furthermore, while the signal in $\langle\sigma_z\rangle$ is sensitive to the recoils of absorbed and reemitted photon, the signal in $\langle\sigma_y\rangle$ is predominantly due to the recoil of the absorbed photon 
in the limit of small geometric phases as it depends on $\phi_{\mathrm{abs}}$ in first and on  $\phi_{\mathrm{em}}$ in second order.

A line profile of the single-photon transition in $^{44}$Ca$^+$ was recovered using the cat state technique (Fig.~\ref{fig:oscillations}b). The $\langle\sigma_y\rangle$-signal amplitude was measured as a function of the spectroscopy laser frequency at a reduced light intensity. The observed line-width of $38 (5)$~MHz is close to the width expected due to the combination of the transition's natural linewidth (22.4~MHz) with the Zeeman splitting caused by an applied magnetic field of 4 Gauss. We wish to emphasise that this profile is due, with high probability, to at most a single infrared photon scattering on an open dipole transition. 

The uncertainty in the spectroscopic signal is determined by quantum projection noise\cite{Itano:1993} in estimating the expectation values of the logic qubit state. The signal-to-noise ratio (SNR) can always be increased by taking more measurements. To compare different methods it is therefore useful to renormalise by the number of measurements $N$ made and consider the sensitivity $\beta=\mathrm{SNR}/\sqrt{N}$ provided by each method.  For comparison we implemented two additional direct detection schemes that do not exploit a non-classical state to amplify the scattering signal. The first and most basic of the two schemes directly measures the probability of the motional state being excited from $\vert 0\rangle$ to $\vert1\rangle$ (phonons) in  a single photon scattering event by applying a red-sideband $\pi$ pulse to the logic ion followed by a measurement of $\langle\sigma_z\rangle$. In the second scheme, $\langle\sigma_y\rangle$ is measured instead. Here, phase sensitive detection of the motional state is enabled by keeping the pulse envelope of the spectroscopy laser phase-locked to the beat signal (at frequency $2\nu$) of the two laser fields manipulating the ions' motion (see Supplementary Section S1).

Table \ref{tab:sensitivity} summarises the experimentally determined sensitivities. Compared to the direct detection method, cat-state spectroscopy using a cat state of $\alpha=2.9(2)$ enabled an 18-fold improvement in measurement sensitivity. Also presented are results showing that, unlike the direct detection methods, cat-state spectroscopy even works \textit{without} ground state cooling. Theoretically, the method should work as well with thermal states that lie within the Lamb-Dicke regime, which can be reached with standard Doppler cooling\cite{Kirchmair:2009}. Experimentally, a loss in contrast is seen whose origin requires additional study. 

In principle, the technique's sensitivity can be further enhanced by making larger cat states. However, experimental errors will limit the useful cat size and thereby the achievable sensitivity. In our experiment the dominant error source is electric field noise on the trap electrodes, which adds another random geometric phase $\phi_{\mathrm{h}}$.  As a result, the contrast of the signal given in equation~(\ref{eq:sigma_z}) needs to be multiplied by a factor $\exp(\overline{-\phi_\mathrm{h}^2}/2)$ with \mbox{$\overline{\phi_\mathrm{h}^2}{=}8R_\mathrm{h}n_{\mathrm{cat}}(\frac{2}{3}\tau_{\mathrm{cat}}+\tau_{\mathrm{wait}})$} where $R_\mathrm{h}$ is the mode's heating rate, $n_{\mathrm{cat}}=|\alpha|^2$ the cat state size in phonons, $\tau_{\mathrm{cat}}$ the time needed for creating the cat state and $\tau_{\mathrm{wait}}$ the time between cat creation and recombination (see \mbox{Methods} and Supplementary Section S4). The larger the cat state, the more detrimental the noise becomes.  For the $\langle\sigma_y\rangle$  measurement in Fig.~\ref{fig:oscillations}a and the experimental parameters $n_{\mathrm{cat}}=8.3(1.0)$, $\tau_{\mathrm{cat}}=50(2)\,\mu$s, $\tau_{\mathrm{wait}}=32\,\mu$s, $R_\mathrm{h}=40(20)$ s$^{-1}$, we expect a contrast reduction factor of 0.91(5) and an overall signal amplitude of $A_{\mathrm{y}}=0.54(4)$, which is close to the observed value.

We have demonstrated a method for detecting light scattering from broad ionic transitions at the single-photon level. Complementing quantum logic spectroscopy for narrow transitions, cat state spectroscopy should provide access to a new range of atomic and molecular ion transitions for precision spectroscopy. The method fundamentally can approach deterministic single photon scattering detection. Assuming that the absorbed and emitted photons have the same wavelength, equation~\eqref{eq:sigma_z} gives the maximum probability of detecting a scattering event, in a single measurement, as 61\%. 
The limit, imposed by the random emission direction of the emitted photon, could be overcome by using multiple vibrational modes or by modifying the emission direction, e.g. by the introduction of a cavity. In our experiment, the achieved sensitivity could be boosted significantly by using an ion trap with a reduced heating rate.

A deterministic measurement of whether an ion has scattered a photon would be of interest to the field of quantum information processing, especially since this measurement could be non-destructive with respect to the joint ion-photon state. 

%%%%%%%%%%%%%%%%%%% Methods section %%%%%%%%%%%%%%%%%%%%%%%%%%

\section*{Methods}
\noindent\textbf{Experimental setup}. Experiments employ a linear Paul trap with a minimal ion-electrode distance of \SI{565}{\micro\meter}. Ions are loaded by isotope-selective photoionization from a resistively heated oven. The bichromatic laser beam used to create and reinterfere the cat state is generated by applying two radio frequencies offset by $2\nu$ to a single acousto-optic modulator, effectively producing a beat-note to which the generator producing the pulse-train of spectroscopy light is locked.\\

\noindent\textbf{Geometric phases contributing to the spectroscopic signal}. 
The total geometric phase that rotates the logic qubit state is given by $\phi{=}\phi_{\mathrm{abs}}{+}\phi_{\mathrm{em}}{+}\phi_\mathrm{h}$ resulting from photon absorption and emission recoil, and random electric fields displacing the ions, respectively. In contrast to the deterministic phase $\phi_{\mathrm{abs}}$ the other two phases are independent random variables with vanishing odd-order moments. This property simplifies the calculation of expectation values at the end of the spectroscopy protocol, leading to $\langle\sigma_z\rangle{=}-\overline{\cos(\phi_{\mathrm{abs}}+\phi_{\mathrm{em}}+\phi_h)}{=}\cos(\phi_{\mathrm{abs}})\overline{\cos\phi}_{\mathrm{em}}\,\overline{\cos\phi_{\mathrm{h}}}$ and  $\langle\sigma_y\rangle{=}\overline{\sin(\phi_{\mathrm{abs}}+\phi_{\mathrm{em}}+\phi_h)}{=}\sin(\phi_{\mathrm{abs}})\overline{\cos\phi}_{\mathrm{em}}\;\overline{\cos\phi_{\mathrm{h}}}$.\\
As spontaneous emission from the $P_{1/2}$ to $S_{1/2}$ state is isotropic, $\overline{\cos\phi}_{\mathrm{em}}{=}\frac{1}{2}\int_0^\pi d\theta \sin\theta\, \cos(\tilde{\phi}_{\mathrm{em}}\cos\theta)$. Writing the integrand as a Taylor series and carrying out the integration results in the sinc-function appearing in equation~(\ref{eq:sigma_z}). \\

Electric field noise at the trap electrodes with frequencies close to the vibrational mode supporting the cat state leads to displacements of the motional state in random directions. White noise acting on a cat of size $n_{\mathrm{cat}}$ for a duration $\tau$ adds a random geometric phase $\phi_\mathrm{h}$ with zero mean and a variance given by $\overline{\phi_\mathrm{h}^2}{=}8R_\mathrm{h}n_{\mathrm{cat}}\tau$~\cite{Turchette:2000a}.  In our model, we assume noise acting on the state throughout the experimental steps of cat state creation, photon scattering and cat recombination, the steps lasting times $\tau_{\mathrm{cat}}$, $\tau_{\mathrm{wait}}$ and $\tau_{\mathrm{cat}}$, respectively.  If a spin-dependent force of constant amplitude is used for cat creation and recombination, we have to make the replacement  $\tau\rightarrow\frac{1}{3}2\tau_{\mathrm{cat}}+\tau_{\mathrm{wait}}$ where the factor $\frac{1}{3}$ accounts for the smaller average cat size during the first and last steps (see Supplementary information S4).  

This work was supported by the European Commission via the integrated project {\sl Atomic QUantum TEchnologies} and a Marie Curie International Incoming Fellowship.

%%%%%%%%%%%%%%%%%%% Bibliography %%%%%%%%%%%%%%%%%%%%%%%%%%
\bibliographystyle{naturemag.bst}

%%%%%%%%%%%%%%%%%%% Figures %%%%%%%%%%%%%%%%%%%%%%%%%%
\newpage

\newpage

\begin{table*}[h]
\caption{\textbf{Comparison of spectroscopic techniques.} 
The sensitivity of various cat state spectroscopy (CSS) techniques using $\alpha = 2.9(2)$, and those which do not employ a non-classical state, are compared (see text). 
Also given are the number of measurements required to reach a confidence of three standard deviations ($\sigma$) in the detection of a scattering event. All errors are calculated from quantum projection noise. (** indicates the result for a CSS measurement \textit{without} ground state cooling.)}

\begin{center}
\begin{tabular}{l | c | r | }
\tableheader{Method}& \tableheader{Sensitivity\\ ($\beta$)} & \tableheader{Measurements\\ to reach $3 \sigma$} \\
\hline
Direct detection $\langle\sigma_z\rangle$ & 0.018(6)\phantom{0} & 2.7(1.9)$\times 10^5$\\
Phase-sensitive $\langle\sigma_y\rangle$ direct detection & 0.107(10) & 7.8(1.5)$\times 10^2$\\
CSS $\langle\sigma_z\rangle$-signal & 0.162(16) & 3.4(0.6)$\times 10^2$\\
CSS $\langle\sigma_y\rangle$-signal & 0.338(16) & 7.9(0.8)$\times 10^1$\\
CSS $\langle\sigma_y\rangle$-signal**& 0.109(14) & 7.6(2.0)$\times 10^2$
\end{tabular}
\end{center}
\label{tab:sensitivity}
\end{table*}%

\clearpage

%%%%%%%%%%%%% Supplementary Information %%%%%%%%%%%%%%%%%%%

\begin{widetext}

\section{{Supplementary information}}

\section{Experimental sequences and definition of signal and noise}
\label{sec:methods}
The following section details the pulse sequences implementing each detection method and explicitly defines the signal $\mu$ and noise $\sigma$. The number of measurements in all cases is $N = \mathrm{min}(N_1, N_2)$, where $N_1$ and $N_2$ are the number of measurements for signal and noise or the oscillation's minimum and maximum, respectively. The following elements are common in all experiments:\\
\textbf{DC:} Doppler cooling, \textbf{OP:} optical pumping, \textbf{SBC:} resolved sideband cooling to initialize the lowest frequency axial vibrational mode of the two-ion crystal in its motional ground state.\\
 Coherent operations (marked in green) have their rotation angle given. \\
The \textbf{hide} - \textbf{unhide} pulses each correspond to a carrier $\pi$ pulse on the logic ion, hiding its electronic $\mathrm{S}_{1/2}$ population in the $\mathrm{D}_{5/2}$ Zeeman manifold to avoid any residual off-resonant scattering during the optical pumping on the $\mathrm{S}_{1/2} \rightarrow \mathrm{P}_{1/2}$ dipole transition of $^{44}$Ca$^+$.\\
The \textbf{detection} in each experiment yields a binary result $x \in [0,1]$, i.e. either the logic ion fluoresces or it does not, which is being decided with respect to a predetermined threshold of counts detected by a photon multiplier tube. The experimental data show that the photon counts measured in the 5~ms detection interval for an ion in the $D_{5/2}$ ($S_{1/2}$) state are well described by Poissonian distributions with average of $12$ ($117$) counts respectively and thus lead to well-separated count distributions. The only deviation from the Poissonian distribution is caused by a decay of the metastable state to $S_{1/2}$ during the detection interval leading to state detection error for an ion in the $D_{5/2}$ state of $0.25\%$. The probability of state detection errors of an ion in the $S_{1/2}$ state is negligible (see also Gerritsma {\sl et al.}, Eur.~Phys.~J.~D {\bf 50}, 13 (2008)). The uncertainty of a measurement average $p$ obtained in $N$ experiments hence is given by quantum projection noise $\Delta p = \sqrt{p (1-p)/ N}$.

\subsection*{Direct detection $\langle\sigma_z\rangle$}

%Doppler cooling\\
%Optical pumping 40\\
%Sideband cooling\\
%Carrier pi pulse (hide logic ion)	\\
%optically pump 44Ca \\
%Carrier pi pulse (unhide logic ion)\\
%Photon tigger (mod 2)\\
%RSB pi pulse\\
%Detection

\begin{figure}[ht!]
\includegraphics[scale=1]{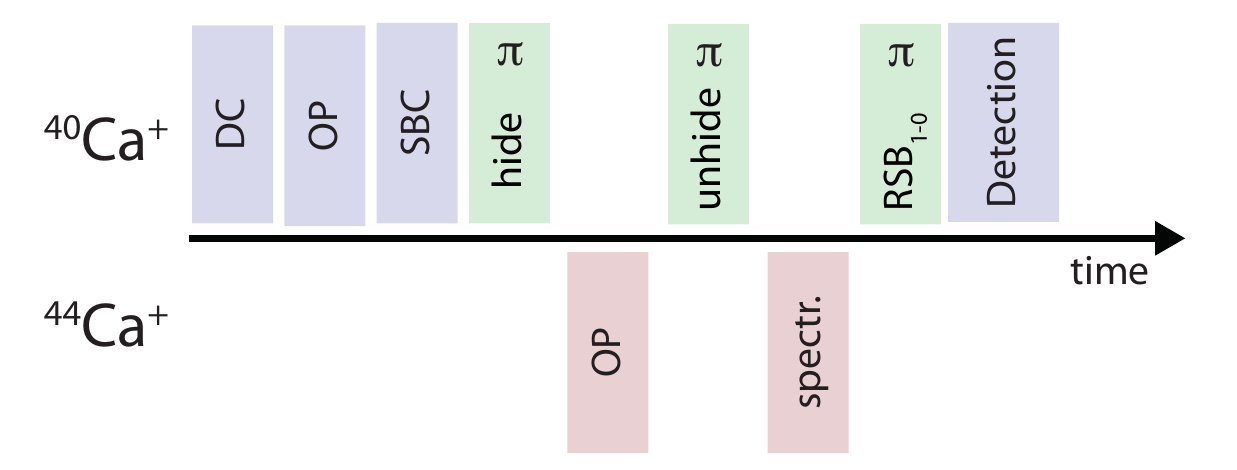}
\label{fig:sequence1}
\caption{\textbf{Pulse sequence used in the direct detection experiment}\newline
Spectroscopy is performed using a single pulse of \SI{10}{\micro\second} duration on the $\mathrm{D}_{3/2} \rightarrow\mathrm{P}_{1/2}$ transition of $^{44}\mathrm{Ca}^{+}$. The red sideband (RSB) pulse prior to detection is calibrated to be a $\pi$ pulse on the $\ket{\mathrm{n}=1} \rightarrow \ket{\mathrm{n}=0}$ motional transition.}
\end{figure}

\noindent The signal $\mu$ is defined as the difference between the averages of $N_1$ experiments taken with the spectroscopy pulse present and $N_2$ experiments without it (data taken in alternating fashion):

$$\mu=\frac{1}{N_1}\sum_{i=1} ^{N_1} \Big(x_{i}^{\mathrm{scatter}}\Big) - \frac{1}{N_2}\sum_{i=1} ^{N_2} \Big(x_{i}^{\mathrm{no\,scatter}}\Big).$$

\noindent Assuming uncorrelated, independent measurements we obtain the noise $\sigma$ as: 
$$\sigma=\sqrt{\Delta x_{\mathrm{scatter}}^2 +\Delta x_{\mathrm{no\,scatter}}^2}.$$

\subsection*{Phase-sensitive $\langle\sigma_y\rangle$ direct detection }

%Doppler cooling\\
%Optical pumping 40\\
%Sideband cooling\\
%Carrier pi pulse (hide logic ion)\\
%optically pump 44Ca \\
%Carrier pi pulse (unhide logic ion)\\
%Photon trigger (always)\\
%switch between delay for fringe min and delay for fringe max\\
%logic ion RSB pi pulse \\
%logic ion pi/2 pulse $\sigma_y$ (0.5 * phase)\\
%Detection

\begin{figure}[ht!]
\includegraphics[scale=1]{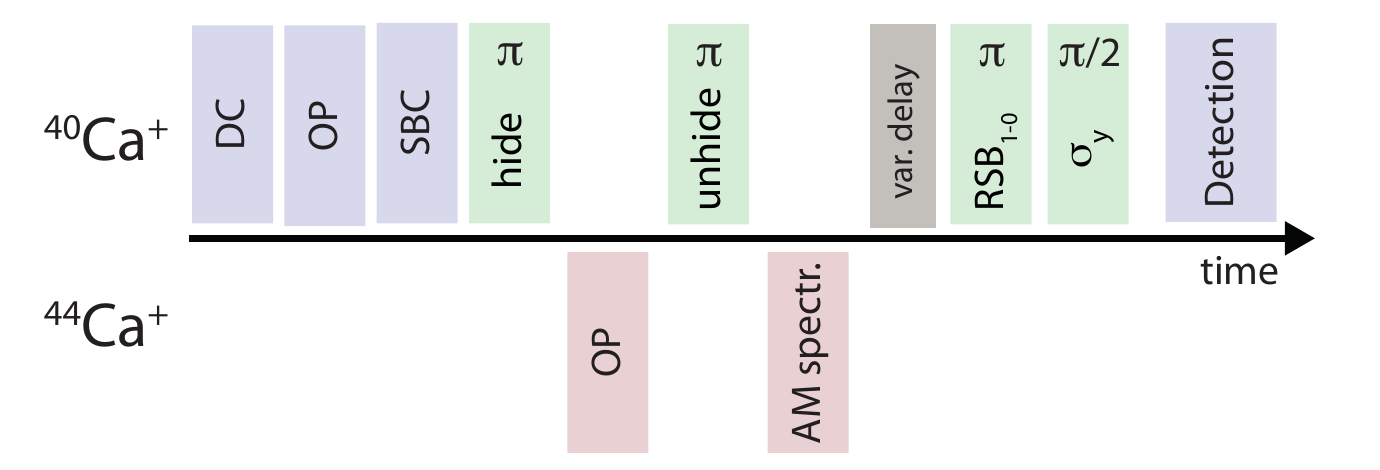}
\label{fig:sequence2}
\caption{
\textbf{Pulse sequence used in the phase sensitive direct detection experiment}\newline
The spectroscopy pulse is amplitude-modulated (AM), yielding short pulses of 60~ns duration that are separated by 800~ns within the \SI{10}{\micro\second} 'envelope'.}
\end{figure}

The idea behind the phase sensitive detection is as follows: Photon scattering provides a momentum kick to the ground state cooled two-ion cystal. We are able to control the time of the scattering event with a temporal resolution far below the ions' oscillation period. This means that after scattering a photon, we know the phase $\phi$ of the coherently displaced motional state, which in the limit of a small displacement can be approximated as $|0\rangle +i\eta e^{i\phi}|1\rangle$. A $\pi$-pulse on the red sideband shifts the phase information to the logic ion by the state mapping $\vert\!\downarrow\rangle(\vert0\rangle +i\eta e^{i\phi}|1\rangle)\longrightarrow (\vert\!\downarrow\rangle +i\eta e^{i\phi}\vert\!\uparrow\rangle)|0\rangle$. The resulting change in the logic ion's electronic state is subsequently detected by measuring the spin projection $\sigma_y$.

The precise control over the timing of the scattering event is achieved by phase-locking the amplitude-modulation frequency to the (virtual) beat-note given by the laser frequencies of the RSB and $\sigma_y$ pulse prior to detection. Varying the time delay between the two pulses and the preceding spectroscopy pulse(s) allows us to recover a fringe pattern representative of the motional phase between the states $\ket{0}$ and $\ket{1}$.

\begin{eqnarray*}
\mu=&\frac{1}{N_1}\sum_{i=1} ^{N_1} \Big(x_{i}^{\mathrm{delay\,max}}\Big) - \frac{1}{N_2}\sum_{i=1} ^{N_2} \Big(x_{i}^{\mathrm{delay\,min}}\Big)\\
\sigma=&\sqrt{\Delta x_{\mathrm{delay\,max}}^2 +\Delta x_{\mathrm{delay\,min}}^2}
\end{eqnarray*}

\subsection*{CSS $\langle\sigma_z\rangle$-signal}

%Doppler cooling\\
%Optical pumping 40\\
%Sideband cooling\\
%Carrier pi pulse (hide logic ion)\\
%optically pump 44Ca \\
%Carrier pi pulse (unhide logic ion)\\
%prepare cat state\\
%Photon trigger (always)\\
%reinterfer cat state\\
%Detection

\begin{figure}[ht!]
\includegraphics[scale=1]{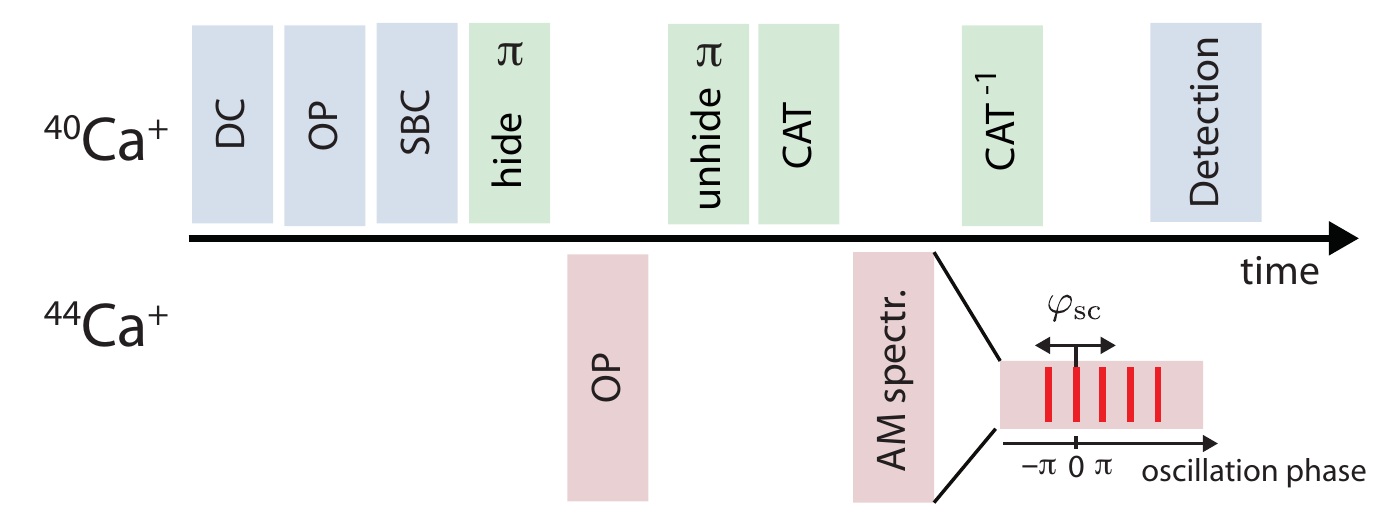}
\label{fig:sequence3}
\caption{
\textbf{Pulse sequence used in the cat state spectroscopy $\langle\sigma_z\rangle$ experiment}\newline
A Schr\"{o}dinger cat state is created (CAT) and reinterfered (CAT$^{-1}$) by a bichromatic laser pulse.}
\end{figure}

\noindent 
The bichromatic light field implementing the coherent displacement operator that produces the cat state (CAT) is  generated using two phase-locked direct digital synthesis (DDS) generators driving a single acousto-optical modulator. It results in two separate diffracted orders set to be resonant with the red and blue sideband of the ion motion, respectively. Both are coupled into the same single mode optical fiber and subsequently delivered to the logic ion, effectively producing a laser beam that is amplitude-modulated at twice the ion oscillation frequency in the trapping potential. This beat note sets the oscillation phase of the cat state in the trap. A third phase-locked DDS generator is now used to produce the amplitude modulation of the spectroscopy pulses, thereby allowing to shift their relative position in time $\varphi_{\mathrm{sc}}$ within the pulse envelope with respect to the cat state oscillation phase.\\

\begin{eqnarray*}
\mu=&\frac{1}{N_1}\sum_{i=1} ^{N_1} \Big(x_{i}^{\mathrm{fringe\,max}}\Big) - \frac{1}{N_2}\sum_{i=1} ^{N_2} \Big(x_{i}^{\mathrm{fringe\,min}}\Big)\\
\sigma=&\sqrt{\Delta x_{\mathrm{fringe\,max}}^2 +\Delta x_{\mathrm{fringe\,min}}^2}
\end{eqnarray*}

%\clearpage
\subsection*{CSS $\langle\sigma_y\rangle$-signal}

%Doppler cooling\\
%Optical pumping 40\\
%Sideband cooling\\
%Carrier pi pulse (hide logic ion)\\
%optically pump 44Ca \\
%Carrier pi pulse (unhide logic ion)\\
%prepare cat state\\
%Photon trigger (always)\\
%reinterfer cat state\\
%logic ion pi/2 pulse $\sigma_y$ (0.5 * phase)\\
%Detection

\begin{figure}[h!]
\includegraphics[scale=1]{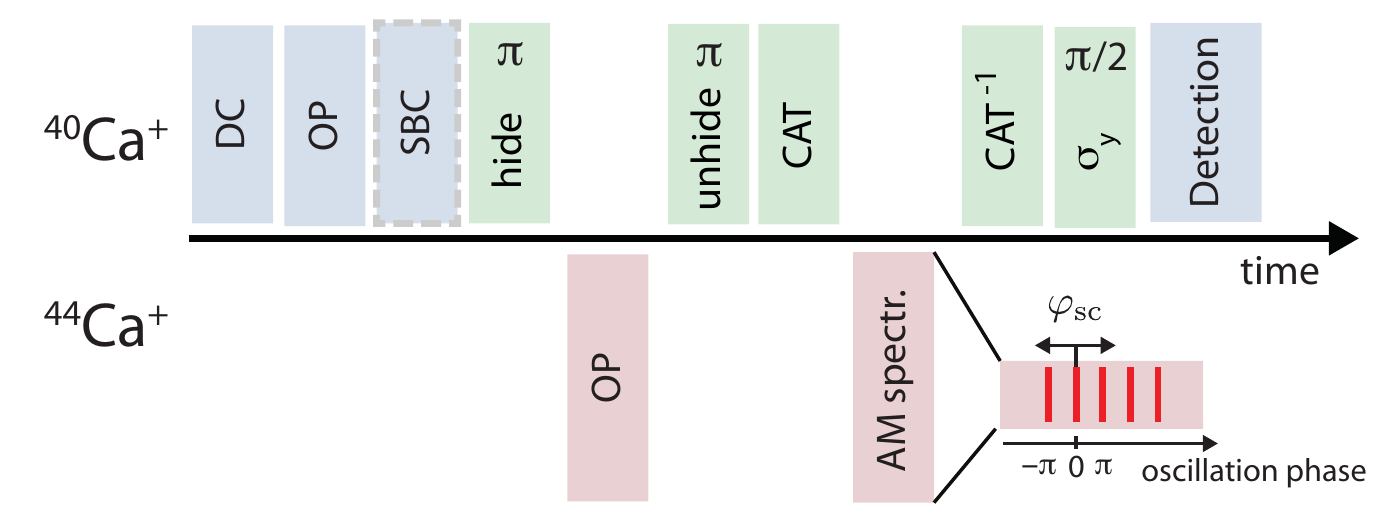}
\label{fig:sequence4}
\caption{
\textbf{Pulse sequence used in the cat state spectroscopy $\langle\sigma_y\rangle$ experiment}\newline
A $\pi/2$-pulse prior to detection rotates the $\sigma_y$-eigenstates into the measurement basis allowing the signal to be detected with increased sensitivity. The high sensitivity allows one to observe a fringe pattern even when ground state cooling (SBC) is omitted.}
\end{figure}

The signal and noise values are obtained in the same way as in the $\langle \sigma_z \rangle$ measurement.

%\clearpage

\section{Timing control over when a photon is scattered}
We use the pulse synchronization for two purposes: first, to improve the visibility of the signal by fixing the geometric phase $\phi$ at its maximum value, and second, to be able to record data in the $\sigma_{y}$ basis. However, even \textit{without} control over the time in which the photon is being scattered, a $\langle\sigma_{z}\rangle$-signal equal to the mean of the fringe pattern in Fig.3a of the main text can still be detected and compared to a control experiment in which no photon has been scattered.

\section{Sensitivity calculation}

Based on the Poissonian detection statistics, the signal-to-noise ratio (SNR) scales as $\displaystyle\mathrm{SNR}=\frac{\mu}{\sigma}\propto\sqrt{N}$, which allows for the definition of a measurement sensitivity $\beta = \mathrm{SNR} /\sqrt{N}$ as a method-independent figure of merit. Its error $\Delta \beta$ can be calculated by error propagation from the individual quantities underlying signal and noise respectively. Based on the obtained signal and noise values, we obtain $\beta$ for each method and calculate the number of measurements (single shots) that have to be averaged to reach a confidence level of $3\,\sigma$ (SNR = 3).
The signal in each method is based on the difference between two quantities (as defined in section \ref{sec:methods} above), which we give in the following as signal A and signal B.

%-- Direct detection --
%scattering signal 		 0.055080(0.001443) - SNR: 38.17 
%background signal 		 0.049421(0.001369) - SNR: 36.09 
%signal 				 0.005659(0.001989) 
%Signal-to-Noise Ratio  		 2.84 
%Exp total 			 25000 + 25050
%Sensitivity beta (from calc SNR) 0.01799
%error sensitivity 		 0.01334
%3 sigma # of experiments (num) 		 27805(41218)

% results below can be found in Mathematica file 'full_calc.nb' in figures_data/OM/certainty_calculation

\begin{table*}[h!]
\caption{\textbf{Results obtained using the different spectroscopic techniques.}}
\begin{center}
\begin{tabular}{l | c | c | c | c | c | c | c| }
\tableheader{Method} &\tableheader{N}&\tableheader{signal A}&\tableheader{signal B}&\tableheader{$|A-B|$}&\tableheader{SNR}& \tableheader{Sensitivity\\ ($\beta$)} & \tableheader{Measurements\\ for $3\sigma$} \\
\hline
(1) & 25000 & 0.055(1) & 0.049(1) & 0.006(2)\phantom{0} & 2.84 & 0.018(6)\phantom{0} & 2.7(1.9)$\times 10^5$\\
(2) & 9850 & 0.446(5) & 0.372(5) & 0.075(7)\phantom{0} & 10.67 & 0.107(10) & 7.8(1.5)$\times 10^2$\\
(3) & 4200 & 0.266(7) & 0.172(6) & 0.094(9)\phantom{0} & 10.51 & 0.162(16) & 3.4(0.6)$\times 10^2$\\
(4) & 4200 & 0.608(8) & 0.376(8) & 0.233(10) & 21.92 & 0.338(16) & 7.9(0.8)$\times 10^1$\\
(5) & 5050 & 0.650(7) & 0.575(7) & 0.075(10) & 7.74 & 0.109(14) & 7.6(2.0)$\times 10^2$
\end{tabular}
\newline
\newline
\end{center}
\label{tab:sensitivity_full}
\end{table*}%

\noindent where the methods are:
\begin{description}
\item[(1)] Direct detection $\langle\sigma_z\rangle$
\item[(2)] Phase-sensitive $\langle\sigma_y\rangle$ direct detection
\item[(3)] CSS $\langle\sigma_z\rangle$-signal
\item[(4)] CSS $\langle\sigma_y\rangle$-signal 
\item[(5)] CSS $\langle\sigma_y\rangle$-signal (without ground state cooling)
\end{description}

\section{Random geometric phases by electric field noise causing motional heating}
Motional heating during the cat state generation, photon scattering and cat state recombination gives rise to additional geometric phases that reduce the contrast of the signal. In this online material we show that due to heating, a random phase $\Phi_h$ is produced during cat state creation and recombination which is given by
\begin{equation}
\langle \Phi_h^2\rangle=\frac{16}{3}n_{cat}R_h \tau     \label{eq:HeatingPhase}
\end{equation}
where $n_{\mathrm{cat}}=|\alpha|^2$ is the size of the cat state in terms of its average phonon number, $R_h$ the motional heating rate, and $\tau$ the length of the two time intervals in which the cat state is created and and subsequently put back again into the initial state. 

To model the heating, we divide the time interval $T=2\tau$ into $N$ small intervals of length $dt=T/N$ and follow the time evolution of one of the motional components of the cat state. In each of these intervals, fluctuating electric fields will displace the motional state by an amount $ d\beta_i=dx_i+dp_i$ where the projections of $d\beta_i$ onto the x- and the p-axis are normally distributed independent random variables with variance $\sigma^2=\frac{1}{2}R_h dt$. This can be seen by noting that, after time $T$, the motional state displacement due to heating alone is given by
\[
\beta(T) = \sum_{i=1}^N d\beta_i = \sum_{i=1}^N dx_i+dp_i
\]
and the average vibrational quantum number becomes
\[
\langle |\beta(T)|^2\rangle=\sum_{i}^N(\langle dx_i^2\rangle+\langle dp_i^2\rangle)=R_h dt N = R_hT
\]
in agreement with the definition of the heating rate.

In a first step, we will study the effect of motional heating on states that are coherently displaced in phase space by some classical well-controlled driving force. In the absence of motional heating, we assume that the initial state $|0\rangle$ is coherently displaced with a (arbitrary) trajectory described by $|\alpha(t)\rangle$. We are interested in the case where the (transient) motional excitation by the coherent drive is much bigger than the excitation by the random fluctuating force causing motional heating. We now look at the effect of a heating event (an undesired kick of the motional state) at time $t_i$ on the state. Before this event, the state was coherently displaced by an amount $\alpha_1=\alpha(t_i)-\alpha(t=0)$ where $\alpha(t=0)=0$. After the event, the coherent displacement is given by $\alpha_2=\alpha(T)-\alpha(t_i)$. The heating event itself displaces the state by an amount $d\beta_i$. In the absence of heating, the final state at time $T$ is given by 
\[
|\alpha(T)\rangle=D(\alpha(T))|0\rangle=e^{-i Im(\alpha_1^\ast\alpha_2)}D(\alpha_2)D(\alpha_1)|0\rangle
\]
where the phase factor cancels the phase arising from the multiplication of the two displacement operations. In the presence of heating at time $t_i$, the state is modified to
\begin{eqnarray}
|\psi\rangle&=&e^{-i Im(\alpha_1^\ast\alpha_2)}D(\alpha_2)D(d\beta_i)D(\alpha_1)|0\rangle\nonumber\\
&=&e^{i Im (\alpha_1^\ast d\beta_i+d\beta_i^\ast\alpha_2)}D(\alpha_2+d\beta_i+\alpha_1)|0\rangle\nonumber\\
&=&e^{i Im (\alpha_1^\ast d\beta_i+d\beta_i^\ast\alpha_2)}|\alpha(T)+d\beta_i\rangle,
\end{eqnarray}
i.e., the state experiences a small additional displacement and is multiplied by the heating-induced random phase
\begin{eqnarray}
\phi_i&=& Im (\alpha_1^\ast\,d\beta_i+d\beta_i^\ast\,\alpha_2)\nonumber\\
&=&  Im (\alpha_1^\ast\,d\beta_i)+Im(d\beta_i^\ast\,\alpha_2)\nonumber\\
&=&  Im (\alpha_1^\ast \,d\beta_i)-Im(d\beta_i\,\alpha_2^\ast)\nonumber\\
&=& Im ((\alpha_1-\alpha_2)^\ast \,d\beta_i). \label{eq:differentialheatingphase}
\end{eqnarray}
If the coherent displacement $\alpha(t)$ is much bigger than the heating-induced displacement, the final state is to a good approximation equal to $|\alpha(T)\rangle$ multiplied by the product of all heating-induced phase factors occuring in the interval $[0, T]$. Equation (\ref{eq:differentialheatingphase}) can be further simplified by splitting $d\beta_i$ into two random variables $du_i$ ($dv_i$) that are oriented orthogonal (parallel) to $(\alpha_1-\alpha_2)^\ast$ and each have variances $\sigma^2$ so that
\[
\phi_i=|\alpha_1-\alpha_2|du_i\sin{\theta_i}
\]
where $\theta_i{=}\pm{\pi/2}$ is the angle between the orthogonal component of $d\beta_i$ and $(\alpha_1-\alpha_2)^\ast$. This term ensures that trajectories $\ket{\alpha(t)}$ in opposite directions pick up opposite geometric phases: as in the case of the two motional components in the generation/recombination of the cat state.  

After simple substitution, the heating-induced geometric phase acquired by the coherent state $\alpha(T)$ is given by
\begin{equation*}
\Phi_g = \sum_{i=1}^N\phi_i=\sum_i|2\alpha(t_i)-\alpha(T_N)-\alpha(0)|du_i\sin{\theta_i}
\end{equation*}
which reduces to
\begin{equation}
\Phi_g =2\sum_i|\alpha(t_i)|du_i\sin{\theta_i} \label{eq:heatingphase}
\end{equation}
in the case  $\alpha(T_N)=\alpha(0)=0$. 

For a cat state that is created and recombined, both components acquire heating-induced phases $\pm\Phi_g$ of opposite sign. Therefore, the relative phase between the two components is  $\Phi_h=2\Phi_g$. We can calculate its mean-squared value:
\begin{equation}
\langle\Phi_h^2\rangle
= 16\sum_{k=1}^N|\alpha(t_k)|^2\langle du_i^2\rangle
= 8R_h\sum_{k=1}^N|\alpha(t_k)|^2\,dt
= 8R_h\int_{0}^T|\alpha(t)|^2\,dt.
\label{eq:meansquaredphase0}
\end{equation}
For 
\[
\alpha(t)=\left\{\begin{array}{lcl}
\sqrt{n_{cat}}\;t/\tau & : & t\le \tau\\
\sqrt{n_{cat}}\;(2\tau-t)/\tau & : & \tau\le t \le 2\tau
\end{array}\right.
\]
the mean squared phase is given by
\begin{equation}
\langle\Phi_h^2\rangle=\frac{16}{3}R_h\tau n_{cat}
\label{eq:meansquaredphase_1}
\end{equation}
which is the result stated at the beginning in eq. (\ref{eq:HeatingPhase}). It is easy to include an additional waiting time $t_{wait}$ between cat creation and recombination in the calculation.
For 
\[
\alpha(t)=\left\{\begin{array}{lcl}
\sqrt{n_{cat}}\;t/\tau & : & t\le \tau\\
\sqrt{n_{cat}} & : &\tau \le t \le \tau+t_{wait}\\
\sqrt{n_{cat}}\;(2\tau+t_{wait}-t)/\tau & : & \tau+t_{wait}\le t \le 2\tau+t_{wait}
\end{array}\right.
\]
integration of eq. (\ref{eq:meansquaredphase0}) yields the mean-squared phase
\begin{equation}
\langle\Phi_h^2\rangle=8R_h n_{cat} (\frac{2\tau}{3}+t_{wait}).\label{eq:meansquaredphase_2}
\end{equation}
For a random variable $X$ that has a Gaussian distribution with $\langle X\rangle = 0$, the equality
\[
\langle\cos X\rangle = \exp(-\frac{\langle X^2\rangle}{2})
\]
holds which enables us to calculate the loss of Ramsey contrast of the cat state spectroscopy from eqs.(\ref{eq:meansquaredphase_1}) or (\ref{eq:meansquaredphase_2}).

\section{Signal dependence on the cat state size}
The Schr\"{o}dinger cat state used for spectroscopy is created by the application of a bichromatic laser pulse resonant with the motional sidebands of the logic ion. The pulse implements the unitary operation \mbox{$U_D = \mathrm{e}^{\alpha(a^\dagger-a)\sigma_x}$}, where $\alpha$ is a coherent displacement, $a$ and $a^\dagger$ are motional creation and annihilation operators, and $\sigma_x$ is a Pauli operator. Applying the operation to the initial state $\ket{\downarrow}_z\ket{0}$, where $\ket{\downarrow}_z$ is the logic ion ground state and $\ket{0}$ the motional ground state of the two-ion crystal, creates a cat state of size $\alpha$:
 $$\frac{1}{\sqrt{2}}\Big(\ket{+}_x - \ket{-}_x\Big)\ket{0} \longrightarrow \frac{1}{\sqrt{2}}\Big(\ket{+}_x \ket{\alpha} - \ket{-}_x\ket{-\alpha}\Big).$$

\noindent In terms of experimental parameters, the size of the cat state $\alpha$ depends on the Lamb Dicke parameter $\eta$, Rabi frequency $\Omega$ related to the light intensity and the duration $t$ of the laser pulse: 
$$\alpha = \eta \Omega t,$$

\noindent which relates to the mean phonon number as ${n_{\mathrm{cat}}}=\alpha^2.$
For the experimental parameters used in the spectroscopy experiment, $\eta = 0.0611$, $\Omega = 2\pi \times \SI{300}{\kilo\hertz}$ (100\% pulse power) and $t = \SI{50}{\micro\second}$, we create a cat state of size $\alpha = 2.88(17)$ or $n_{\mathrm{cat}}=8.3(1.0)$. Here, the Rabi frequency $\Omega$ is defined such that a pulse of duration $\tau=2\pi/\Omega$ on the carrier transition results in a $2\pi$ rotation of the Bloch vector. The Lamb-Dicke parameter $\eta$ describes the reduction of coupling strength on the blue sideband transition (as compared to the carrier transition) when the $^{40}$Ca$^+$ ion is excited on the ground-state cooled COM-like mode of the mixed two-ion crystal. Uncertainties in the determination of the cat size are due to errors in the calibration of the Rabi frequency and the pulse area due to intensity pulse shaping.\\
The following data sets investigate the scaling of the signal size $A_y$ with the duration (Figure \ref{fig:time_oscillations}) and the optical power (Figure \ref{fig:power_oscillations}) of the bichromatic laser pulse creating the cat state. 

\begin{figure*}[ht!]
\centering
\includegraphics[scale=0.45]{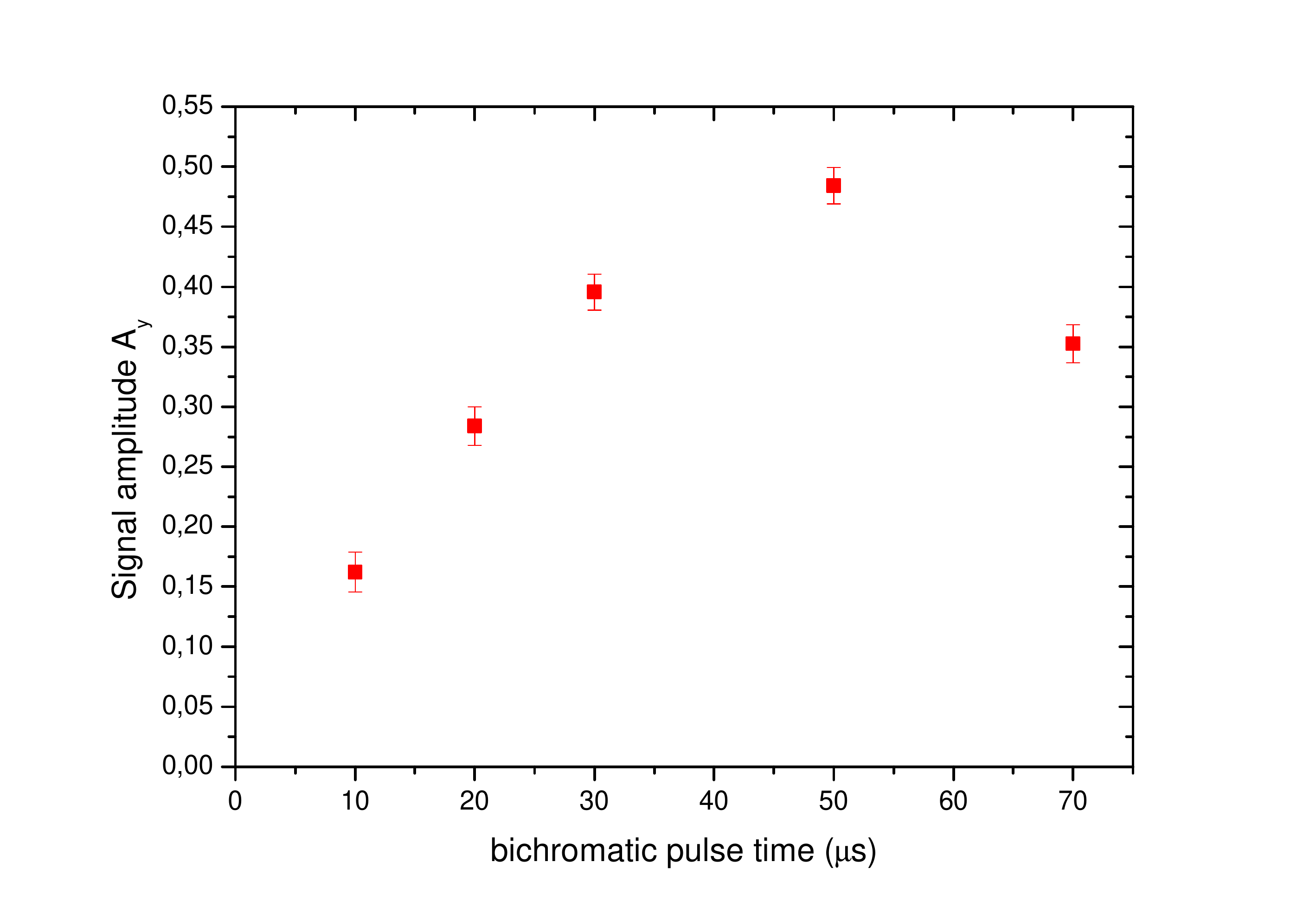}
\caption{\textbf{Signal size as a function of bichromatic pulse time}\newline
For larger cat sizes decoherence (mostly due to heating in our setup) sets in which led us to limit the experiment to the cat state size of $\alpha = 2.9$ produced by the \SI{50}{\micro\second}-pulse. (Note that the maximum signal size in this data set is smaller due to a drift in the trap frequency at the time of data taking.)}
\label{fig:time_oscillations}
\end{figure*}

\begin{figure*}[ht!]
\centering
\includegraphics[scale=0.45]{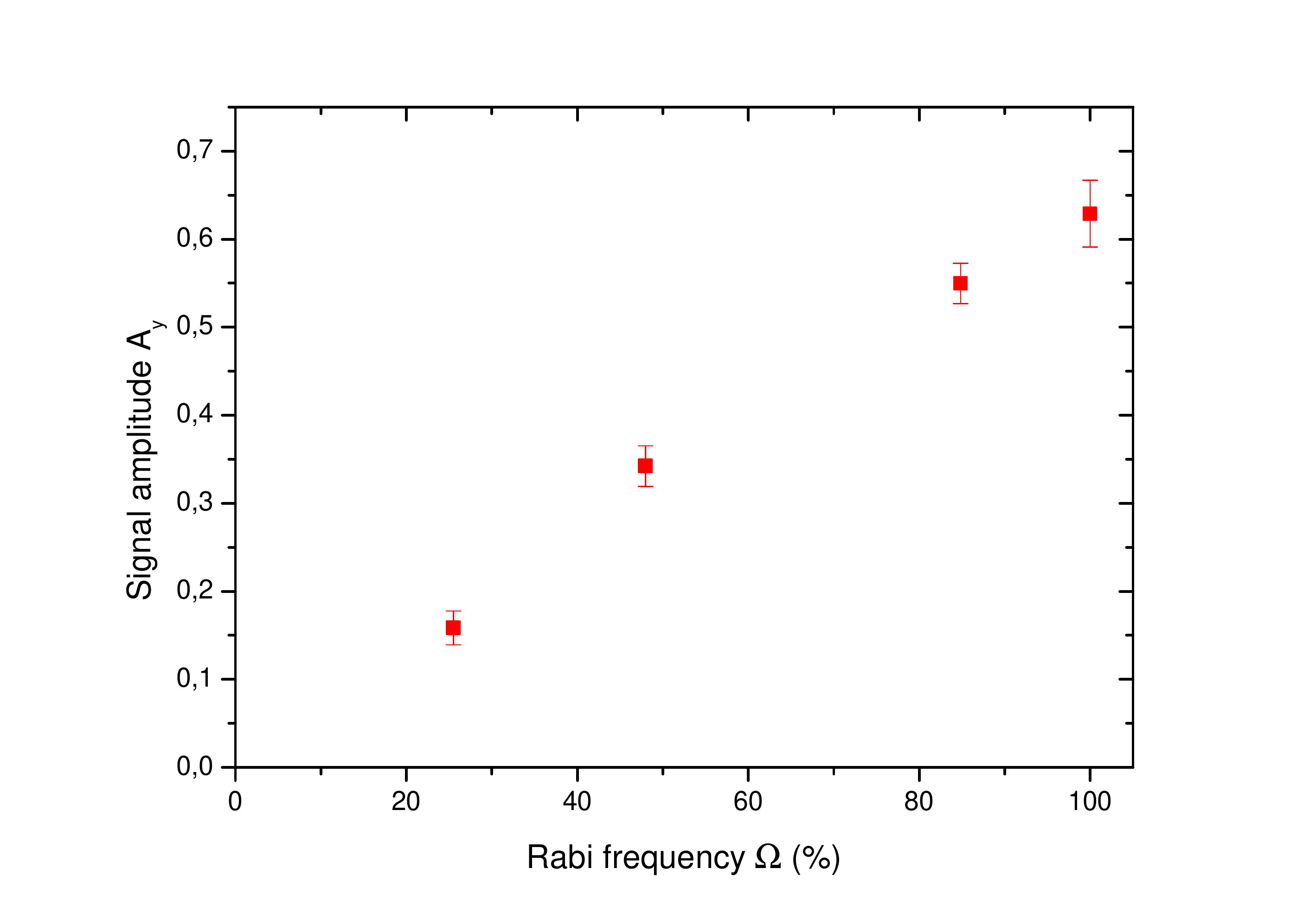}
\caption{\textbf{Signal size as a function of bichromatic pulse power} \newline Using a pulse duration of $t=\SI{50}{\micro\second}$, the graph shows the amplitude $A_y$ obtained from a weighted sinusoidal fit to the data as we vary the optical power ($\propto \Omega^2$) in the bichromatic laser beam creating the cat state.} 
\label{fig:power_oscillations}
\end{figure*}

\section{Spectroscopic line profile at different laser powers}
In order to reduce power broadening of the $D_{3/2}\rightarrow P_{1/2}$ in $^{44}Ca^{+}$, we recorded the spectrum depicted in Fig.~3b of the main text for three different laser powers (in an interleaved fashion). All three traces including weighted Gaussian fits can be seen in Figure \ref{fig:line_profiles} below.\newline
The transition's natural line-width of 22.4~MHz appears broader in our case due to the Zeeman splitting caused by an applied magnetic field of 4.1~Gauss. 
For clarity, the main text only shows the trace for \SI{7}{\micro\watt}. Here, the probability of scattering one or more photons in a single experiment is less than 1 such that, even on resonance, some experiments do not contribute a signal, leading to the reduced amplitude.

\begin{figure*}[ht!]
\centering
\includegraphics[scale=0.45]{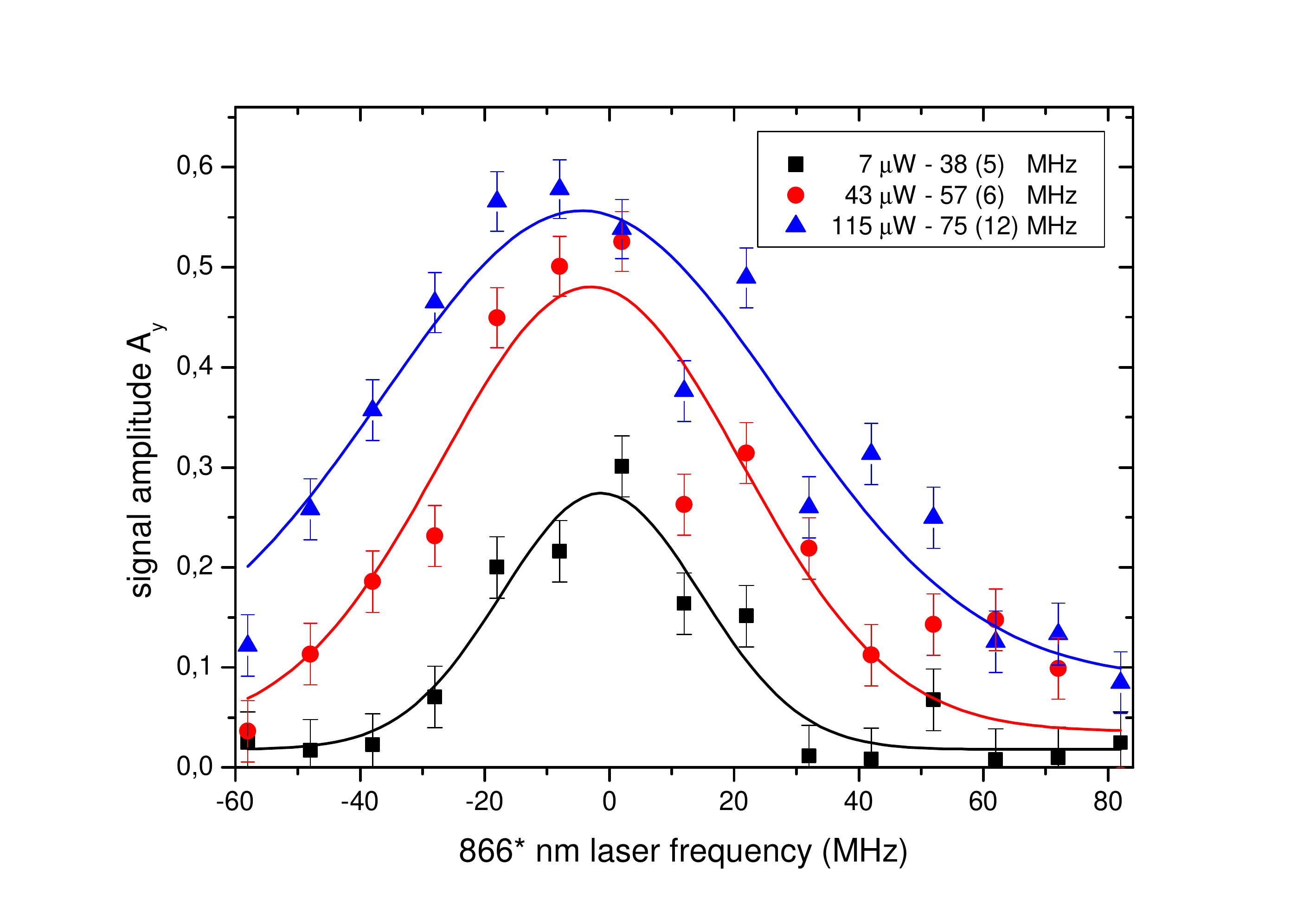}
\caption{\textbf{Line profile for different spectroscopy laser powers} The error bars are derived from quantum projection noise and enter the Gaussian fits as weights. The line widths are given as full width at half maximum values.} 
\label{fig:line_profiles}
\end{figure*}

\end{widetext}

\end{document}